# GOVERNMENT LICENSE

# Opportunities for Retrieval and Tool Augmented Large Language Models in Scientific Facilities

Michael H. Prince[1], Henry Chan[2], Aikaterini Vriza[2], Tao Zhou[2], Varuni K. Sastry[3], Matthew T. Dearing[4], Ross J. Harder[1], Rama K. Vasudevan[5], Mathew J. Cherukara[1,*]

[1] Advanced Photon Source, Argonne National Laboratory, IL, USA
[2] Center for Nanoscale Materials, Argonne National Laboratory, IL, USA
[3] Argonne Leadership Computing Facility, Argonne National Laboratory, IL, USA
[4] Business and Information Systems, Argonne National Laboratory, IL, USA
[5] Center for Nanophase Materials, Oak Ridge National Laboratory, Oak Ridge, TN, USA

## Abstract

Upgrades to advanced scientific user facilities such as next-generation x-ray light sources, nanoscience centers, and neutron facilities are revolutionizing our understanding of materials across the spectrum of the physical sciences, from life sciences to microelectronics. However, these facility and instrument upgrades come with a significant increase in complexity. Driven by more exacting scientific needs, instruments and experiments become more intricate each year. This increased operational complexity makes it ever more challenging for domain scientists to design experiments that effectively leverage the capabilities of and operate on these advanced instruments. Large language models (LLMs) can perform complex information retrieval, assist in knowledge-intensive tasks across applications, and provide guidance on tool usage. Using x-ray light sources, leadership computing, and nanoscience centers as representative examples, we describe preliminary experiments with a Context-Aware Language Model for Science (CALMS) to assist scientists with instrument operations and complex experimentation. With the ability to retrieve relevant information from facility documentation, CALMS can answer simple questions on scientific capabilities and other operational procedures. With the ability to interface with software tools and experimental hardware, CALMS can conversationally operate scientific instruments. By making information more accessible and acting on user needs, LLMs could expand and diversify scientific facilities' users and accelerate scientific output.

---

* mcherukara@anl.gov

# 1. Introduction

Conversational agents such as ChatGPT built on large language models (LLMs) are upending various industries and professions[1]. Their impact spans education[2,3], talent acquisition[4], and many other sectors. A report by McKinsey and Company estimates that generative AI, including LLMs, could add trillions of dollars annually to worldwide productivity[5].

LLMs also have the potential to revolutionize every aspect of scientific discovery, including accelerating literature search, aiding in experimental design, providing data summaries, and publication writing and editing[6]. LLMs have also been shown to be fast learners, requiring only a few training examples to quickly gain domain expertise, e.g., for materials property prediction and inverse design[7].

At scientific user facilities, such agents could significantly contribute to nearly all facets of operating advanced scientific instruments and conducting experiments. These agents could help new users navigate different aspects of the facility, such as the proposal submission process, safety policies, and operating practices. They could also help users design suitable experiments to address their scientific inquiries, including selecting the appropriate instrument, determining the necessary modalities, and advising on the sample preparation. Additionally, during the experiment, they could guide basic instrument operation. Despite their potential, LLMs suffer from several dangerous limitations, including the tendency to 'hallucinate' answers when faced with questions not part of their training.

We describe preliminary experiments with retrieval[8] and tool-augmented[9] LLMs tailored for scientists' needs. Scientific context-aware LLMs leverage existing data stores to augment their capabilities, allowing them to perform better in domains that are not part of the pre-training data, thereby eliminating the need to perform any fine-tuning on the domain-specific information that can be computationally expensive and time-consuming. In this work, we describe using LLMs to create an agent with knowledge of scientific user facilities. We describe the initial results of assisting scientists through basic experiment design, instrument operation, and direct experimental tool control.

**1.1 Related work**

LLMs have demonstrated strong performance as autonomous agents across multiple science domains[10]. The key components that make the LLM function as the agent's brain are memory, planning, and tool use. Recently, Yager discussed the importance of memory by building domain-specific chatbots for physical science research. Their demonstration involves the creation of a context window using text chunks from scientific documents in PDF format. When the LLM is composing its reply, it uses this context to formulate the answer[11].

ChemCrow is another example of a domain-specific LLM agent designed to accomplish tasks related to materials design, organic synthesis, and drug discovery by integrating several chemistry tools. Their workflow combines planning via Chain of Thought reasoning (CoT) and tool utilization[12]. In another recent work, researchers built an Intelligent Agent based on GPT, capable of searching hardware documentation and controlling multi-instrument systems to conduct experiments. An important consideration in this type of research is the evaluation of the chatbots or LLMs for scientific applications. In this context, Zaki et al. developed a question-answering dataset for measuring the performance of GPT-3.5 and GPT-4. However, the performance is measured without considering the improvement when an LLM is provided with additional domain-specific scientific context[13]. Jablonka et al. described results from a hackathon applying LLMs to various scientific tasks.[14]

## 2. Early Experiments

### 2.1 Overview of CALMS

The core of CALMS consists of four components: a large language model (LLM)[15], conversational history that allows follow-on queries, semantic search over document stores to retrieve the most relevant context given a question from the user[16], and instrument tools that the LLM can use when instructed to by the user[17]. Figure 1 shows a schematic of the structure of CALMS. When provided with appropriate context and tools, CALMs can either answer complex technical questions that the LLM has no prior knowledge of from its training[18] or execute an experiment or perform a computation using the appropriate tools[17]. We describe each component in detail in the following sections. We compare responses from two state-of-the-art LLMs, OpenAI's GPT-3.5 Turbo and an open-source model Vicuna, over questions related to experimental design, operation, and ability to drive an instrument successfully. We note that the CALMS framework is independent of the LLM, and other open-source or closed-source LLMs can be swapped without changing the framework.

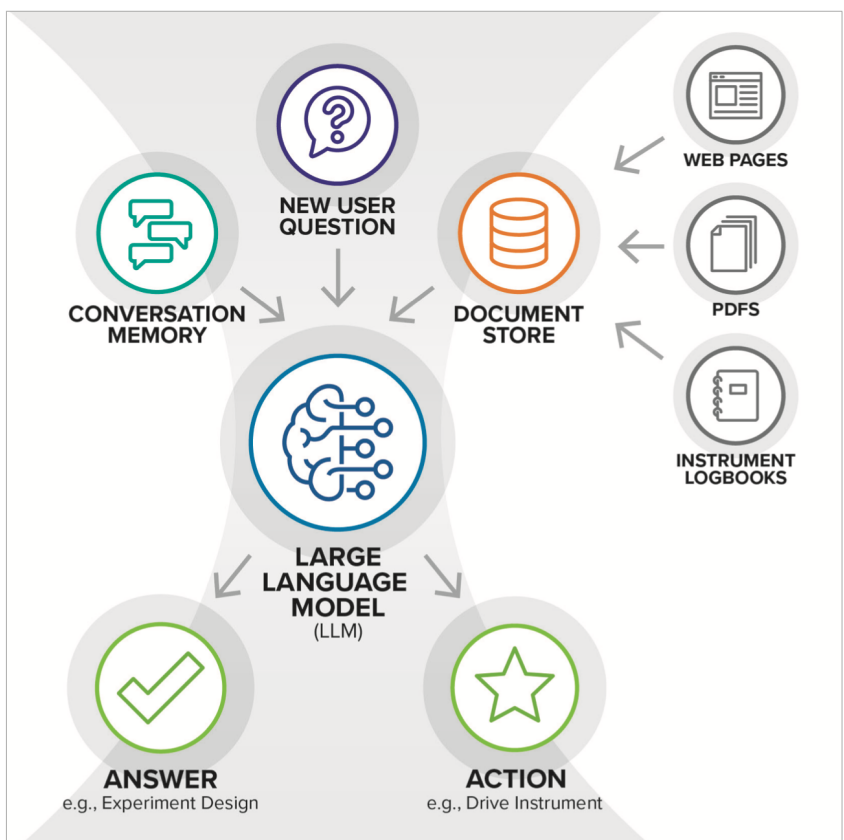

*Figure 1: Overview of CALMS: CALMS uses a large language model in conjunction with conversational memory, document stores, and experimental tools to answer user queries or take action to drive an instrument.*

**2.1.1 GPT-3.5**

The Generative Pretrained Transformer 3.5 (GPT-3.5) is a near state-of-the-art large language model developed by OpenAI and released in late 2022. Similar to its predecessors and successor, GPT-4, the model architecture is based on the Transformer[19], which has demonstrated effectiveness in various natural language processing tasks, including generating text, answering questions, writing content with human-like creativity, and even generating computer programming code. These outcomes result from predictive inference based on an input sequence of text processed through the language patterns the model learned from a vast training set.

The most performant of the *pre*-GPT-4 class of models, GPT-3.5 Turbo, was trained with 175 billion parameters and is available at the time of this writing with context window sizes of 4,096 or 16,384 tokens. The context window corresponds to the maximum number of tokens (i.e., words, punctuation, or even sub-strings of words) the model can process during a single inference procedure. The content included in this window represents a simple type of memory that can be leveraged to provide background as effective guidance toward predicting subsequent words. This feature can significantly impact the quality and relevancy of the model output and has led to the burgeoning field of "prompt engineering" to develop and implement best practices.

While the model is proprietary, Microsoft makes it available through the Azure cloud with an enterprise subscription. Hosting of the model resides on a dedicated and restricted instance, so data remains local, is not retained by default, and is not shared with third parties. This approach alleviates some challenges with using public models, such as ChatGPT from OpenAI.

The interaction with GPT-3.5 starts with an API call authenticated using a unique key. This key is essential for accessing OpenAI's capabilities securely and ensuring that only authorized users can make requests. Before making the API call, specific parameters, i.e., temperature and top_p, guide the model's response generation. The temperature parameter controls the randomness of the output and takes values between 0-1. A higher temperature leads to more creative output, while a lower temperature results in more deterministic outputs. The top_k parameter limits the model's choices to the 'k' most likely next words at each step of the generation process. The top_p parameter is an alternative to the temperature that focuses on a subset of predictions that cumulatively add up to the probability 'p'. After these parameters are set and the API call is made, the model can be used either for chat completion or embeddings generation. For chat completion tasks, the model generates a continuation of the provided text input based on the context and parameter set. For the embeddings generation task, the model processes the input text to create a vector representation that captures the semantic meaning of the text. These embeddings can be used for various applications like text similarity assessment or semantic search.

**2.1.2 Open-source LLMs**

LLMs have transformed the AI field in recent years with the advent of large and complex models like GPT-3 and GPT-4. Though their performance and capabilities are unparalleled, they often require very large compute resources, thus limiting the number of individuals and enterprises that can develop and maintain systems of this scale. Moreover, with limited or no transparency about the model specifications, data, and development methods, these "closed" source LLMs raise concerns regarding adherence to AI ethical standards (including fairness, morality, accountability, and misuse, among others). Open-source LLMs alleviate these concerns by democratizing the AI community with free and unrestricted accessibility to models and information driven by platforms and communities with a transparent and collaborative approach. Open-source LLMs also provide the advantage of being accessible to modification to tailor it to a specific task. HuggingFace is such a resource that provides a platform where researchers, academics, and data scientists can share datasets.[20] HuggingFace also provides the necessary tools within their Transformers library to fine-tune and deploy ML workflows for open-source models, including LLMs. With a wide range of modalities of models and data available, it also has the added advantage of seamlessly switching and testing between different models with minimal code changes.

HuggingFace also provides the "Open LLM Leaderboard evaluation suite" that evaluates and compares the performance of different LLMs based on a relative Elo rating strategy. Traditionally, the Elo rating system has been a method used in the field of chess to calculate the relative skill levels of players based on the outcome of the games. A similar strategy is applied to the model responses, where human annotators rate responses for helpfulness and truthfulness in a pairwise setting.[21] At the time of finalizing the paper, "lmsys/vicuna-13b-v1.5" was ranked among the top five pertained models with an average score of 61.63. This model was obtained by fine-tuning Llama 2 on supervised instruction downstream tasks trained on conversations collected from ShareGPT with a sequence length of 2048.[22]

In the following sections, we compare responses from Vicuna 1.5-16k against GPT-3.5 across a series of tasks. Our evaluation encompasses tasks including general comprehension of user questions, domain-specific context retrieval, quality of the answers, and utilization of tools.

### 2.1.3 Context retrieval through semantic search

One of the prominent use cases of LLMs is context retrieval through semantic search, which is a process that involves extracting relevant information from a large text body based on the meaning and context of the query instead of relying on keyword matching. Currently, Claude has the largest context window of 100k tokens, corresponding to around 75,000 words.

Two approaches are possible to extend the context retrieval capabilities of models beyond their training corpus. First, fine-tuning the model with domain-specific training data requires the careful curation of data and often necessitates significant up-front investment either in local GPU resources (e.g., Llama was trained on 2048 GPUs)[23] in the case of open-source or in cloud compute costs in the case of closed-source models. A great deal of expertise in language modeling, distributed training, transformers, and data curation is also typically required to fine-tune these models successfully. However, this approach will likely provide the best results when a sufficiently large corpus of data for fine-tuning is available.

A second approach to enhance the capabilities of LLMs is to leave the model weights unchanged but provide appropriate additional context to help answer user queries. This strategy is typically cost-efficient, and a plethora of frameworks now exist to enable facile interaction of the LLM with document stores. A remaining challenge, however, is that even the largest models, e.g., GPT-4-128k, can only accept input text up to 128k tokens, corresponding to approximately 300 pages of text. While future models might allow even larger input sizes, the current limitations prevent passing in all relevant documentation[24]. To address this gap, available documentation is pre-processed for context retrieval, which is performed via a lightweight embedding model. All technical documents are split into token windows of a pre-determined size, and their embeddings are stored in a vector database, e.g., ChromaDB, collection[25]. When a question is sent, the user's text is embedded, and the top $N = 4$ text chunks closest to the user input are retrieved from the store. These context chunks are then included in the prompt provided to the conversational LLM.

### 2.1.4 Conversational memory

At their core, LLMs are probabilistic models that learn sentence or paragraph completion. In other words, given a sequence of words, they are trained to predict the most likely completions of that sentence or paragraph. Models that have been fine-tuned for chat have an additional training step that teaches them how to complete conversations, either through an unsupervised approach by learning from human-provided conversations or, more popularly, through reinforcement learning with human feedback (RLHF).[26] Such fine-tuned chat models can complete human-like conversations, and we leverage this ability by providing the conversational history as part of the subsequent prompt. As previously discussed, due to the limited amount of text that can be provided in the prompt, reproducing the entire conversational history is often not possible. In essence, the problem is how to provide the LLM with short-term memory to answer questions more effectively when faced with a limited context window. Several strategies have been proposed to address this; one option that CALMS adopted is to use a moving window over the conversation history to include only the last $K = 6$ conversations between the user and the LLM. Supplementary Figure 1A shows the prompt to the LLM with context and memory.

### 2.1.5 Software and experimental tools

To further augment the use of LLMs, current research is exploring if the LLMs can go beyond text and reason about the physical world. For example, work from Google and Microsoft have demonstrated using NLP to achieve robotic tasks.[27,28] This process involves building a series of design principles that include high-level robot APIs or function libraries, special prompting structures, and human feedback via text.

Building upon this strategy, the current implementation of CALMS incorporates capabilities such as calling the Materials Project API[29], an instrument control software (spec$^{TM}$)[30], and is based on Chain-of-Thought prompting and the ReAct framework.[31,32] These functions are provided as a list of tool names, including the description of the way they interact with scientific equipment and details about the user inputs and outputs. The tool is then instructed to answer a user-provided prompt by leveraging the available tools. A parser interprets the LLM output for tool call requests. If a tool is called, then the parser inserts the output into the prompt, and the model continues generating actions until a response is provided. Supplementary Figure 1B shows the prompt to the LLM with the tool description.

In effect, using conversational memory, context-retrieval, and specified tools creates a dynamic pre-prompt that allows the LLM to parse and respond to user input with a detailed understanding of the context behind those queries.

## 2.2 AI-assisted experimental design

We first explore the ability of context-aware LLMs, such as those implemented in CALMS, to respond to highly technical questions typical of those encountered by new scientific facility users. CALMS is queried with questions concerning user facilities, including the Advanced Photon Source (APS), Center for Nanoscale Materials (CNM), Argonne Leadership Computing Facility (ALCF), and Center for Nanophase Materials Sciences (CNMS). A summary of responses with Vicuna and GPT-3.5 Turbo with and without context retrieval is shown in Figure 2. The AI response is first graded by their helpfulness (denoted Rel in Fig. 2 for relevance), 'yes' for giving a relevant answer to the question, and 'no' for an irrelevant answer. Next, their truthfulness is evaluated by checking if hallucinations (denoted Hal) appear in the response. While we appreciate that relevance and hallucinations are not correlated, a relevant response can be hallucinated, and an irrelevant response can be entirely truth-based. Finally, the completeness (denoted Com) of the AI response is graded with a score of 0–5. Details on the scoring are included in the Methods section. A complete list of the queries and AI responses is found in the data availability section, with highlighted key differences. We caution that the reader might observe slightly different results because the LLMs outputs are non-deterministic.

| Experimental Design | | | | | | | | | | | | | |
|---|---|---|---|---|---|---|---|---|---|---|---|---|---|
| Test Cases | | CNM query | | | ALCF query | | | APS query | | | CNMS query | | |
| Criteria | | Rel | Hal | Com | Rel | Hal | Com | Rel | Hal | Com | Rel | Hal | Com |
| w/o ctx | Vicuna | Yes | Yes | 0 | N/A | N/A | 0 | Yes | Yes | 0 | No | Yes | 0 |
| | GPT-3.5 | Yes | Yes | 0 | N/A | N/A | 0 | Yes | Yes | 3 | No | Yes | 0 |
| w/ ctx | Vicuna | Yes | No | 4 | Yes | No | 2 | Yes | Yes | 1 | Yes | No | 4 |
| | GPT-3.5 | Yes | No | 5 | Yes | No | 5 | Yes | No | 4 | Yes | No | 5 |

*Figure 2: Scoring of experimental design questions with and without context. We score the models on relevance, absence of hallucination, and completeness of response.*

Given an accurate context, CALMS can always provide relevant answers to questions specific to each user facility. Rarely was hallucination observed. In general, GPT-3.5 Turbo tends to provide a more complete answer than Vicuna, scoring multiple 5s in the completeness grading. A score of 5 indicates that the users are provided with all the necessary information to select the appropriate tool to complete their task.

Without appropriate context, the language models have a strong tendency to hallucinate. As a representative example, we queried the models on the name of the automated image simulation framework developed at the CNM. Both Vicuna and GPT-3.5 Turbo provided the user with a made-up tool called ImageSim (an apparent portmanteau for image and simulation) (see Figure 3). We note that both models, with context, were able to retrieve the correct answer, which is a tool called ingrained.[33] Without context, irrelevant answers have been observed, and occasionally, the models decline to answer the question, citing insufficient knowledge (see for example Supplementary Figure 2). Supplementary Figures 3 and 4 show further examples illustrating the importance of context-aware responses.

**User Input**

Are there any tools developed at the CNM related to automated framework for image simulation?

**AI Response (without context)**

Vicuna: Yes, one such tool is called "CNM-ImageSim". It was developed by researchers at the Center for Nanoscale Materials (CNM) to simulate images with atomic resolution. This tool uses density functional theory (DFT) calculations to predict the electronic structure and properties of materials, which can be used to generate realistic images of these materials.

GPT-3.5: Yes, there is a tool developed at CNM called ImageSim. It is an automated framework for image simulation that allows users to create realistic synthetic images for a wide variety of applications, including computer vision, machine learning, and robotics. The tool incorporates advanced algorithms and models to simulate various physical phenomena that affect image acquisition, such as noise, blur, distortion, and occlusion. It also includes a user-friendly interface that enables users to easily specify the key parameters of the simulation process, such as the lighting conditions, the scene geometry, and the camera properties. Would you like more information on how ImageSim works and its features?

**AI Response (with context)**

Vicuna: Yes, one example of a tool developed at the CNM related to automated frameworks for image simulation is the "ingrained" framework mentioned in the paper published in Small. This framework allows for the fusion of atomic-resolution materials imaging simulations into corresponding experimental images.

GPT-3.5: Yes, the Center for Nanoscale Materials (CNM) has developed an automated framework for image simulation called "Ingrained". It uses density functional theory calculations and computer vision to simulate and match scanning tunneling microscopy images. The tool has been published in a Small report and is open-source. Its main use is to fuse simulations of atomic resolution images of materials into the experimental images to which they correspond.

*Figure 3: Demonstration of AI-assisted experimental design specific to the CNM user facility: (top) initial question from a user, (bottom-left) response from AI without context, (bottom-right) response from AI with context. Major differences between the responses are highlighted (in yellow) to illustrate the importance of context.*

## 2.3 AI-assisted experiments

We further queried the models' ability to provide operational assistance on advanced scientific and computational resources at the abovementioned user facilities. A summary of responses with Vicuna and GPT-3.5 Turbo with and without context retrieval is shown in Figure 4. Similar behavior was observed compared to what was noted for the experimental design queries. Without context, both Vicuna and GPT-3.5 Turbo tend to hallucinate. Even when they are being truthful (i.e., not hallucinating), the answers they provide are typically not directly helpful.

| Test Cases | | CNM query | | | ALCF query | | | APS query | | | CNMS query | | |
|---|---|---|---|---|---|---|---|---|---|---|---|---|---|
| Criteria | | Rel | Hal | Com | Rel | Hal | Com | Rel | Hal | Com | Rel | Hal | Com |
| w/o ctx | Vicuna | Yes | Yes | 0 | No | Yes | 0 | No | No | 0 | No | Yes | 0 |
| | GPT-3.5 | Yes | Yes | 0 | No | Yes | 0 | Yes | No | 1 | No | Yes | 0 |
| w/ ctx | Vicuna | Yes | No | 4 | Yes | No | 3 | Yes | No | 3 | Yes | Yes | 3 |
| | GPT-3.5 | Yes | No | 3 | Yes | No | 5 | Yes | No | 5 | Yes | No | 5 |

*Figure 4: Scoring of experimental operations assistance questions with and without context. As before, we score the models on relevance, absence of hallucination, and completeness of response.*

For instance, when queried about the procedure to start a tomographic scan, Vicuna, without context, responded with a truthful answer concerning the computational tomography (CT) scan in a hospital, which is unrelated to the tomographic scan performed at the APS (Figure 5). In this case, the truthful yet unhelpful answer Vicuna provided was due to the LLM mistaking the tool being queried with a different subject well-known in the public domain. GTP-3.5 Turbo realized the nuances in the question and suggested the user be more specific about the type of tomographic scan they are inquiring about. It then provided a general answer about the need to submit a user proposal to perform a tomography experiment.

**User Input**

How do I start a tomography scan?

**AI Response (without context)**

**Vicuna**: To start a tomography scan, you will need to follow these steps: 1) Prepare the patient by ensuring they are in a comfortable position and have removed any metal objects that may interfere with the scan. 2) Position the patient on the scanner bed so that the area of interest is aligned with the opening of the scanner. 3) Adjust the settings on the machine according to the type of scan being performed and the patient's size. 4) Begin the scan by pressing the start button on the machine. It is important to note that only trained medical professionals should perform tomography scans as they require specialized knowledge and equipment.

**GPT-3.5**: Are you referring to a specific type of tomography scan? Different types of tomography scans may have different requirements for starting the procedure. Please provide more context so I can give you a more accurate answer.
**Human**: How do I start a tomography scan at the APS?
**GPT-3.5**: To start a tomography scan at the APS (Advanced Photon Source), you will first need to submit a proposal to the APS beamline staff detailing your research objectives and the type of experiment you will be conducting. You will then need to schedule a beamtime slot and prepare your sample for the scan according to the specific instructions provided by the beamline staff. Once your sample is ready, you will need to control the beamline instrument and collect the data using specialized software. The beamline staff can provide training on how to use the software and operate the instrument. Do you have any other questions about the process?

**AI Response (with context)**

**Vicuna**: You can start a tomography scan using the tomoscan command. This command is used to manage tomography scans and is part of the Micro-CT software stack.

**GPT-3.5**: To start a tomography scan, you can enter the acquisition parameters in the tomoScan screen and press the green "Start Scan" button. Alternatively, you can run a single scan from the command line interface with the command "tomoscan single". If you want to do a vertical, horizontal, or mosaic scan, you can use the commands "tomoscan vertical", "tomoscan horizontal", or "tomoscan mosaic" respectively. You can find more details about each of these commands in the tomoScan documentation.

*Figure 5: Demonstration of AI-powered operational assistance specific to the APS user facility: (top) initial query from a user, (bottom-left) response from AI without context, (bottom-right) response from AI with context. Major differences between the responses are highlighted (in yellow) to illustrate the importance of context.*

However, with context, both Vicuna and GPT-3.5 Turbo provide consistently relevant and truthful answers, with GTP-3.5 Turbo again faring better on the score of completeness than Vicuna. Still using the tomographic scan procedure query as an example, Vicuna responded with a short answer regarding the use of the tomoscan command, while GPT-3.5 Turbo's answer included solutions with the graphic user and the command line interface with a reference to the relevant documentation (Figure 5). Supplementary Figures 5, 6, and 7 show further instances of the importance of context-aware responses.

**2.4 Automated execution**

Through the ReAct framework[32] and API calls, the CALMS framework could be configured to drive experiments in an agentic manner. To demonstrate this capability, a test was conducted on a real-world diffractometer simulating a common user operation.

For a user to take a diffraction measurement, an area detector, usually mounted to a robotic arm, must be moved into a set position relative to the sample and incoming x-ray beam. This angle is determined by the intensity of the beam, the material being observed, and the peak position selected by the user. To do this process manually, the user must first look up lattice information via a materials database. Then, the user must copy over the lattice information and the selected peak into the beamline interface, controlled by SPEC, to move the motors into position.

To automate this process, two LLM calls were implemented: GetLatticeConstants and SetDiffractometmer. For the LLM to successfully perform the action, it must first input the material into a call that queries the Materials Project API to return the lattice constants to the model. Next, it must combine the output of this call with the peak positions from the user input to execute a second call. Figure 6 describes in detail the sequence and parameters of the calls. In the example shown, a user requests the diffractometer to be set to the 012 Bragg peak of $WSe_2$. Within the CALMS framework, the LLM parses the query and recognizes the material and Bragg peak. It then uses a tool to call The Materials Project[29] to obtain the lattice information for WSe2. It then passes that information to a second tool that uses the Spec software to calculate the motor positions and then moves the diffractometer to that position.

These calls were implemented via LangChain's "Structured input ReAct" framework. This library API allows developers to define the parameters and types for multi-parameter input and requires the model to use valid JSON to perform a call. In our tests, we observed that the open-source models could not consistently follow this syntax. All results shown were run with OpenAI's GPT-3.5 model. We could execute this sequence of commands via a single user prompt on beamline 34 at the APS. A video of the commands being run on the beamline is provided in the supplemental material of this paper and can also be viewed online.[34]

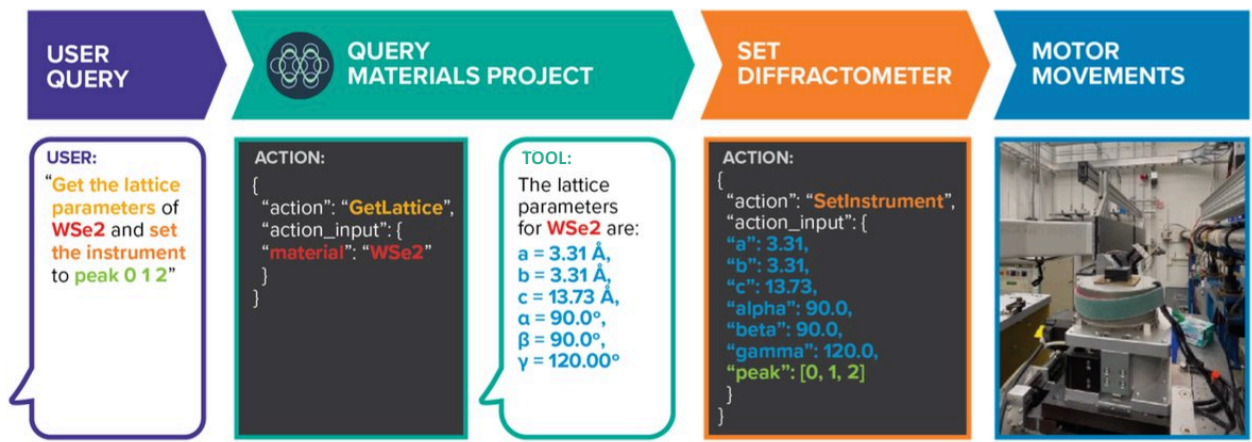

*Figure 6: Workflow demonstrating the execution of a robot motor move based on a user query to CALMS. CALMS parses the user query to extract the material and Bragg peak, queries Materials Project for the lattice constants, and then calculates the position of and moves the beamline diffractometer to the Bragg peak requested by the user. A video of the demonstration is available at https://danielzt12.github.io/latest_news/2023/11/20/operating-scientific-instruments-with-LLMs.html*

## 3. Summary and Outlook

In summary, we described the development of a Context-Aware Language Model for Science (CALMS) and exemplified the promising potential of leveraging LLMs for accelerating scientific experiments and discovery. Preliminary results obtained using CALMS backed by GPT-3.5 and Vicuna illustrate the applicability of LLMs in experimental design and operation guidance, leveraging the zero-shot capability of LLMs via prompt engineering[35]. The results emphasize the importance of context in addressing the typical issue of hallucination in LLMs. CALMS powered by GPT-3.5 and Vicuna both performed reasonably well in the Q&A tasks, but the results reveal limitations in the open-source models compared to state-of-the-art closed-source models such as GPT-3.5 Turbo. This gap is particularly pronounced in tool usage, where we could not execute the experimental workflow with the open-source model Vicuna. With the investment and extraordinary rate of progress in open-source models and fine-tuning methods, we expect the performance gap to the closed-source models to narrow.

Traditionally, model performance over the years has shown power law scaling with an increase in the number of model parameters, dataset size, and compute budget. However, emerging research demonstrates that it is equally important to scale the training size with an increase in model parameter size to obtain optimal training.[36] Additionally, with instruction fine-tuning and RLHF, we see that the even smaller models of the size 1.3B parameters can equal the performance of the 175B GPT-3-base model.[37] With several innovations like SELF-Instruct[38] (that can enhance the model performance by 33% over the GPT3 base model with fine tuning on the self-generated instruction without any human intervention) and Chain-of-Thought prompting[15,39], the broad consensus is that these performance gains will continue to push the state of the art. With such anticipated advanced model performance, we envision future capabilities could include leveraging decades of information recorded in e-logs as well as the ability to extract code or commands and execute experimental workflows fully autonomously.

## 4. Methods

**Scoring methodology:** 1/1 pt is automatically given if the response is rated 'yes' on relevance and 'no' on hallucination. 2/2 pts is given if the instructions can be executed by the user without error, 1/2 if minor errors are present that are easily fixable, and 0/2 if major errors are present in the response. 2/2 pts is assigned if the user is given all the information they need for their request, 1/2 if minor details are missing, and 0/2 if crucial details are missing that affect the execution. The maximum score is 5 pts.

## 5. Code and Data Availability

All code, models, and data will be made publicly available on GitHub with the published manuscript.

## 6. Acknowledgments

Work performed at the Center for Nanoscale Materials and Advanced Photon Source, both U.S. Department of Energy Office of Science User Facilities, was supported by the U.S. DOE, Office of Basic Energy Sciences, under Contract No. DE-AC02-06CH11357. This research used resources of the Argonne Leadership Computing Facility, a U.S. Department of Energy (DOE) Office of Science user facility at Argonne National Laboratory and is based on research supported by the U.S. DOE Office of Science-Advanced Scientific Computing Research Program, under Contract No. DE-AC02-06CH11357. M.J.C and H.C also acknowledge support from the U.S. Department of Energy, Office of Science, Office of Basic Energy Sciences Data, Artificial Intelligence, and Machine Learning at DOE Scientific User Facilities program under Award Number 34532.